\documentclass[reprint,preprintnumbers, amsmath,amssymb, aps]{revtex4-2}

\usepackage{graphicx}
\usepackage{dcolumn}
\usepackage{bm}
\usepackage{changes}
\usepackage{color}

\begin{document}



\title{Universal Law of Coiling for  a Short Elastic Strip Contacting Within a Tube}
\author{ Jeng Yi Lee,$^{1,*}$ Hao-Yu Lu,$^{2}$ and Ray-Kuang Lee$^{2,3,4,5}$}
\affiliation{$^{1}$Department of Opto-Electronic Engineering, National Dong Hwa University, Hualien 974301, Taiwan\\
$^{2}$Department of Physics, National Tsing Hua University, Hsinchu 30013, Taiwan\\
$^{3}$Institute of Photonics Technologies, National Tsing Hua University, Hsinchu 30013, Taiwan\\
$^{4}$Physics Division, National center for theoretical sciences, Taipei, 10617, Taiwan\\
$^{5}$Center for Quantum Technology, Hsinchu 30013, Taiwan}

\date{\today}

\begin{abstract} 
We find that there exists a universal law of coiling not only for a long elastic strip contacting within a tube but also for a short one. 
Here the elastic strip we consider has the ratio of   $2 < L/R \le 2\pi$ for its length $L$ to the tube radius $R$.
By varying the ratio of $L/R$, we identify four types of deformation for such a short elastic strip, namely, two point-contact, three point-contact, continuous-contact, and self-contact.
With theoretical formulas in closed forms and experimental demonstration, these four types are  verified  for any elastic strips contacting within a tube, irrespective of elastic properties, strip lengths, and tube radius.
Our results on coiling can be readily applied to a variety of physical systems, including thin flexible electronic devices, van der Waals materials in scroll shape, and DNA packaging into viral capsids. 
\end{abstract}

\maketitle

\noindent {\it Introduction.}\textemdash
Packing a long wire, fiber, or strip  inside a container happens in many different systems, such as folding elastic wire in a spherical cavity~\cite{wire5, wire6,wire7,wire8,wire9,wire10,wire11,wire12}, bending graphene sheets or van der Waals material  in scroll shapes~\cite{graphene1,graphene2, graphene3,graphene4}, packing DNA into viral capsids~\cite{DNA1,DNA2,DNA3,DNA4,DNA5,DNA6,DNA7,DNA8,DNA9}, and curling sheets in confirmed structures~\cite{sheet1,sheet2,sheet3}.

By considering a long elastic strip inside a smooth and solid cylindrical tube,  a universal law of coiling was discovered~\cite{nature1,nature2}.
Irrespective of the tube size, the total strip length, and the elastic bending stiffness, when an elastic strip intrinsically flat  is coiled inside a tube, the innermost strip is detached from the tube wall at multiple-layered curls.
With free-of-friction contact forces between strip-strip and strip-tube interfaces, the tangential angle of  a detached strip at the free end to the tube's tangent is $24.1^{\circ}$; while the opening angle, subtended from the detached region, is  $125.2^{\circ}$~\cite{nature1,nature2}.

 This universal phenomenon was derived in theory first, then has been observed on a variety of surprisingly different length scales and in unexpected systems, including mechanical, biological, and condensed matters.
Although experimental measurements and theoretical analyses suggest that the classical elastic plate model deviates to describe the bending deformation of monolayer graphene due to absence of  in-plane $\sigma-$ bonding, the continuum plate phenomenology can still be well employed in  glued multilayers due to the mediation of van der Waals force \cite{graphene1,graphene2,graphene3,graphene4}.

Instead of long strips, in this {\it Letter}, we reveal that there also exists a universal law of coiling  for a short elastic strip contacting within a tube.
Here, we refer to a short strip as one with a ratio of the strip length $L$ to the tube radius $R$ between $2 < L/R \le 2\pi$.
With the help of Kirchhoff's equations for a elastica, we first theoretically identify four types of deformation for such a short elastic strip in Fig.~1, labelled, (a)  two point-contact, (b) three point-contact, (c) continuous-contact, and (d) self-contact, which are characterized for arbitrary elastic materials.
In experiments,  see the illustrations in Figs.~1(e)-(h), we fabricate samples in polyvinyl chloride (PVC) and  polyethylene terephthalate (PET) with different lengths and thickness to verify our theoretical prediction, resulting in good agreement.
As similar scenarios of such a short strip can be easily found in a wide range of physical systems, our results on coiling a short elastic strip can be readily applied to thin flexible electronic devices,  photovoltaic solar cells, van der Waals materials, and DNA packaging.\\

\begin{figure*}[htp]
\centering
\includegraphics[width=1\textwidth]{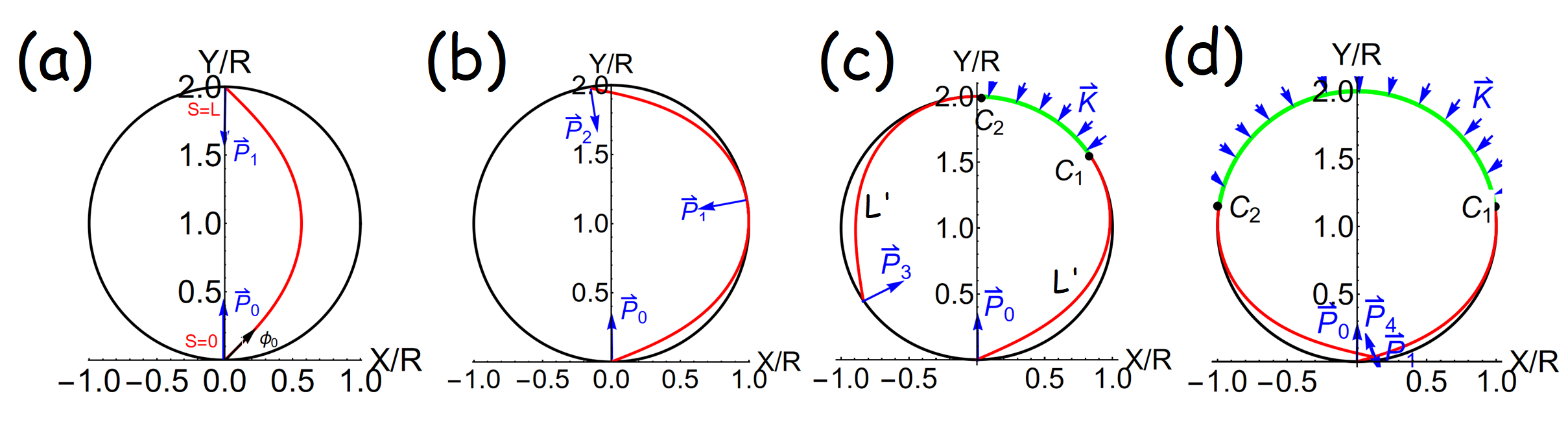}
\includegraphics[width=1\textwidth]{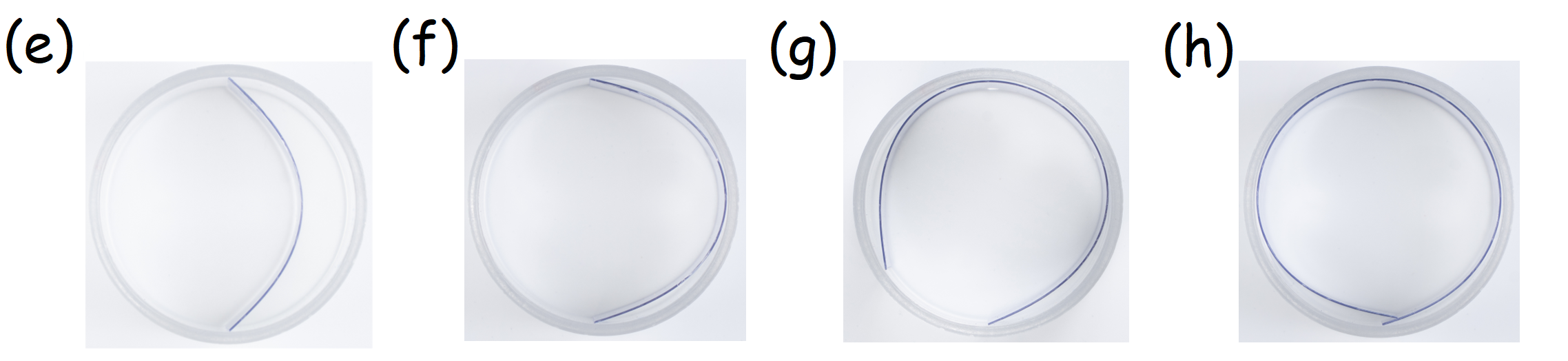}
\caption{Four different types of deformations are identified as the universal law of coiling,  when the ratio of the strip length $L$ to the tube radius $R$ is  $2 < L/R \le 2\pi$.
The First row (a)-(d) shows the simulation results for  (a) $L/R = 2.37$,  (b) $3.2$, (c) $5.1$, and (d) $2\pi$, respectively. Here the red curves denote the detached strips; while the green curves denote the continuous contact parts of the strips.
The Second row (e)-(h) shows the corresponding experimental measurements  as a comparison (see Table 1 for more details).
Here, the four types of deformations are (a, e) two point-contact, (b, f) three point-contact,  (c, g) continuous-contact, and (d, h) self-contact.
Related force analysis diagrams are also depicted in (a-d), denoted in blue colors.} 
\end{figure*}

\noindent {\it Theory: Kirchhoff's equations.}\textemdash  
As illustrated in Figs. 1(a-d), we model an intrinsically flat strip as  a quasi one-dimensional elastica. 
Based on Kirchhoff's theory, the force and moment equations in the static equilibrium read~\cite{book1,book2},
\begin{eqnarray}
&&\frac{d\, \vec{F}(s)}{ds}+\vec{K}(s)=0,\\
&&\frac{d\, \vec{M}(s)}{ds}+\hat{t}(s)\times\vec{F}(s)=0.
\end{eqnarray}
Here, $s$ is the arc-length along the elastic strip, $\hat{t}(s)$ is the unit tangent vector, $\vec{K}(s)$ is the external force per unit length, $\vec{F}(s)$ and $\vec{M}(s)$ denote resultant stress force and moment at $s$, respectively. 
In our theoretical analysis, we assume the elastica  free of external bending moment, i.e., without friction  between strip-strip and strip-tube interfaces and free of gravity.
In case of a planar deformation, we have $\vec{M}=B\, \hat{t}\times d\hat{t}/ds$. 
Here  $B$ is the bending stiffness composed by $B=Y\, I$, with  $Y$ being Young's modulus and $I$ being the moment of inertia, which is equivalent to the quadratic form of curvature in the bending free energy \cite{book2}.
Based on Eqs.~(1-2), we start our analyses by increasing the stripe length ratio $L/R$: from slightly larger than $2$ to $2\pi$ and reveal the emergence of four different types of deformation.\\



\noindent {\it Two point-contact.}\textemdash   
First of all, we consider the strip length slightly larger than $2R$, resulting in only two point-contacts upon the strip at $s=0$ and $s=L$, as the numerical calculation shown in Fig.~1(a), as well as  the corresponding experimental illustration  in Fig.~1(e).

Now, the associated point-contact forces at $s=0$ and $s=L$ are $\vec{P}_0$ and $\vec{P}_1$, respectively, as blue arrows in Fig.~1(a).
Due to absence of friction,  the directions of external forces is normal to the tube wall. 
Moreover, in static equilibrium, we have $\vec{P}_0=-\vec{P}_1$. 
 With $\vec{K}=0$, by Eq.~(1), the stress force $\vec{F}(s)$ is constant throughout the strip, i.e., $\vec{F}(s)=-\vec{P}_0$.
By resorting to moment balance of Eq.~(2), one can obtain a curvature equation,
\begin{equation}\label{1}
\frac{d\phi(s)}{ds}=\sqrt{\frac{2P_0}{B}(\sin\phi-\sin\phi_0)}.
\end{equation}
Here, $P_0=\vert \vec{P}_0\vert$ and $\phi(s)$ is the tangential angle with respect to X-axis. Moreover, we denote $\phi_0\equiv \phi(s=0)$, and use a zero  moment condition at $s=0$, corresponding to $d\phi/ds\vert_{s=0}=0$.
As the length of the strip projected onto the Y-axis is always $2R$ and the strip length $L$ is conserved, we thus have two geometric constraints:
\begin{eqnarray}\label{2}
R&=&\int_{\phi_0}^{\frac{\pi}{2}}\frac{\sin\phi}{\sqrt{\frac{2P_0}{B}(\sin\phi-\sin\phi_0)}}\, d\phi,\\
\frac{L}{2}&=&\int_{\phi_0}^{\frac{\pi}{2}}\frac{1}{\sqrt{\frac{2P_0}{B}(\sin\phi-\sin\phi_0)}}\, d\phi,
\end{eqnarray}
in which two unknowns $\phi_0$ and $P_0$ are involved. Then, by eliminating $P_0$,  consequently $\phi_0$ can be calculated  for a given value of  $L/R$, irrespective of bending stiffness $B$, but under the crucial geometric constraint for the strip at $s=L/2$, i.e., $0 \leq X(s=L/2)/R < 1$. 

Accordingly,  we numerically find the  condition  to support $\phi_0$ and reach the constraint $0.358< \phi_0\leq \pi/2$, which gives us the corresponding strip length in this two point-contact region: 
\begin{eqnarray}
L/R \in [2,3.033); \, \text{two point-contact region}.
\end{eqnarray}
We want to remark that this result holds for any elastic strip systems.
As we increase $L/R$, the strip bends more, reflecting a monotonous decrease of $\phi_0$ as shown in Fig.~2(a).
Furthermore, with Eqs.~(4-5), one can also introduce  a dimensionless force:
\begin{equation}\label{pointforce}
\frac{P_0R^2}{B}=\frac{1}{2}[\int_{\phi_0}^{\frac{\pi}{2}}\frac{\sin\phi}{\sqrt{\sin\phi-\sin\phi_0}}d\phi]^2,
\end{equation}
which is a function of $\phi_0$. 
In Fig.~2(b), we display this dimensionless force $P_0R^2/B$ as a function of $L/R$.
In this two point-contact region, although this dimensionless force $P_0R^2/B$  monotonously decreases as  $L/R$ increases, when with material property and stripe length ($B$ and $L$) are fixed,  the associated point-contact force  $P_0$ increases for a short radius $R$.
The value of $P_0R^2/B=2.47$ at $L/R=2$ corresponds to the critical force for the classical Euler's buckling bifurcation with $B\pi^2/L^2$ \cite{book2,book3,buckling}.
\\ 

\noindent {\it Three point-contact.}\textemdash   When the middle segment of the strip ($s=L/2$) makes a point contact with the tube, a three point-contact situation occurs, as illustrated from numerical calculation and experimental measurement in in Figs.~1(b) and 1(f), respectively.
Here, as shown in Fig.~1(b), we denote  three external point-contact forces upon the strip as $\vec{P}_0$, $\vec{P}_1$, and $\vec{P}_2$.
In the static equilibrium, their vectorial sum is zero.

By applying  mirror symmetry to the strip, we conclude that the magnitudes of these external forces at two free ends are identical $\vert \vec{P}_0\vert =\vert \vec{P}_2\vert=P_0$, although their direction is different.
Again, in the detached region from $s=0$ and $s=L/2$, due to  absence of external forces we also have $\vec{F}(s)=-\vec{P}_0.$
Then, by employing the geometric constraint for point-contact and that the strip length $L$ is conserved, as well as the zero  moment at free ends, two unknowns can be determined:  the tangential angles with respect to X-axis at $s=0$ and $s=L/2$, i.e., $\phi_0$ and $\phi_1 \equiv \phi(s=L/2)$ can be determined through the following conditions,
\begin{eqnarray}\label{position1}
R\sin\phi_1&=&\int_{\phi_0}^{\phi_1}\frac{\cos\phi }{\sqrt{\frac{2P_0}{B}[\sin\phi-\sin\phi_0]}}\, d\phi,\\
R-R\cos\phi_1&=&\int_{\phi_0}^{\phi_1}\frac{\sin\phi }{\sqrt{\frac{2P_0}{B}[\sin\phi-\sin\phi_0]}}\, d\phi.
\end{eqnarray} 
Again, by eliminating the unknown $P_0$ for a given value of  $L/R$, we can sufficiently determine $\phi_0$ and $\phi_1$, independently of  $B$. The detailed derivation is given in Supplementary Materials.

By requiring the radius of curvature at $s=L/2$ being $R$, corresponding to the onset of continuous-contact, we can find out the supported values of $\phi_0=0.358=20.51^{\circ}$, which defines the maximum strip length obeying the three point-contact situation:
\begin{equation}
L/R \in [3.033 ,4.176 ); \,\text{three point-contact region}.
\end{equation}
The corresponding dimensionless force $P_0R^2/B$ for the three point-contact case is
 \begin{equation}\label{p0}
 \frac{P_0R^2}{B}=\frac{2(\sin\phi_1-\sin\phi_0)}{\sin^2\phi_1},
 \end{equation}
which has an implicit dependence on $L/R$.
When $L/R=3.033$, the emergent point-contact at the middle strip makes $\vec{P}_2$ have a horizontal component in order to obey  static equilibrium.
In this way, the strength of $\vec{P}_2$ along the vertical direction would be decreased, leading to reduce $\vert \vec{P}_0\vert$ in Fig.~2 (b).
As a result, when the length ratio $L/R > 3.033$ the tangential angle  $\phi_0$ increases, as shown in the inset of Fig.~2 (a).\\

 \begin{figure}[t]
\centering
\includegraphics[width=0.5\textwidth]{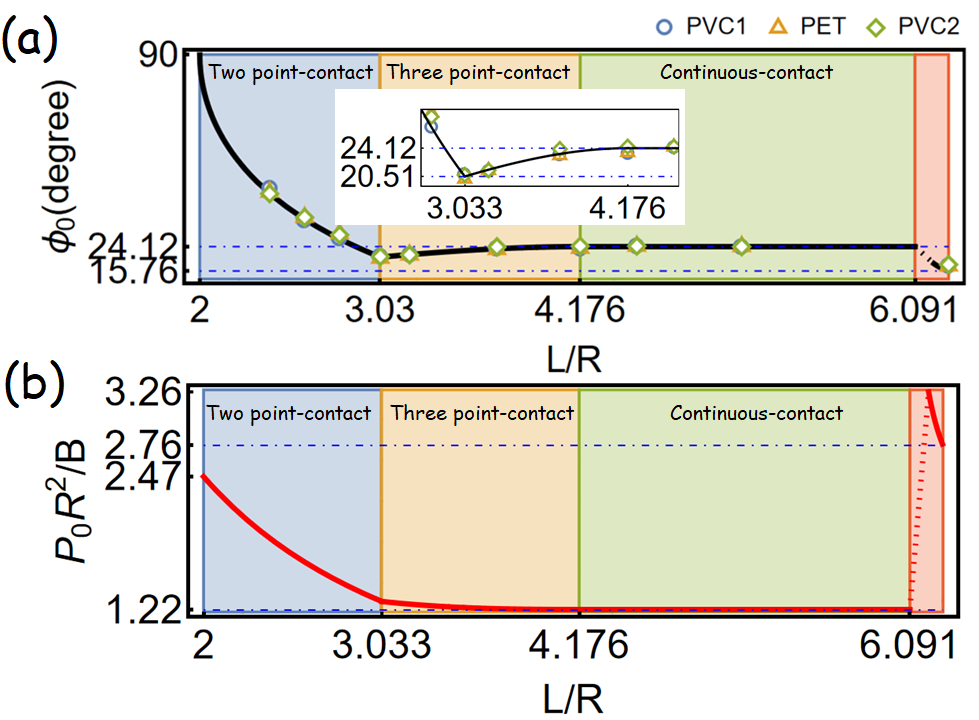}
\caption{(a) In terms of the length ratio $L/R$, we identify different deformation regions through numerical simulations  for the tangential angle $\phi_0$ (black curve), and  three sets of experimental measurements:  $\circ$ ($0.5$mm thickness of PVC, denoted as PVC1), $\diamond$ ($0.1$mm thickness of PVC,  denoted as PVC2), and $\triangle$ ($0.2$mm thickness of PET, denoted as PET).
(b) The corresponding dimensionless force $ P_0R^2/B$ is also depicted as a function of the length ratio $L/R$.} 
\end{figure}

\noindent {\it Continuous-contact.}\textemdash   
 When $d\phi/ds\vert_{s=L/2}$ becomes $1/R$, the strip deformation switches to  continuous contact, as illustrated in 
in Figs.~1(c) and 1(g) from numerical and experimental results, respectively.
Now, we have a finite region making   continuous contact with the tube, as marked by the green colored curve in Fig.~1(c).
The associated pressure distribution $\vec{K}$ exerted by the tube remains constant in magnitude \cite{note2}.

We mark the two ends of this continuous contact as $C_1$ and $C_2$.
By considering the geometric constraint of the curvature at $C_1$ being $1/R$, as well as zero moment at the free end $s=0$, again one can have the following two equations  involving two unknown tangential angles at $s=0$ and $s=L^\prime$,
\begin{eqnarray}\label{univ3}
2\sin\phi_0&=&\sin\phi_1,\\
\frac{\sin\phi_1}{1-\cos\phi_1}&=&\frac{\int_{\phi_0}^{\phi_1}\frac{\cos\phi}{\sqrt{(\sin\phi-\sin\phi_0)}}\, d\phi}{\int_{\phi_0}^{\phi_1}\frac{\sin\phi}{\sqrt{(\sin\phi-\sin\phi_0)}}\,d\phi}.
\end{eqnarray}
For  details see  Supplementary Materials.

By numerical calculation, we obtain that the tangential angle at the free end is $\phi_0=0.421=24.12^{\circ}$; while the opening angle is  $\phi_1=2.185=125.2^{\circ}$.
Interestingly, these two angles are exactly the same angles for a long stripe~\cite{nature1,nature2}. 
The curvature equation for this detached strip, as well as that boundary conditions, return to the same  scenario as that of multiple layered curls.
We indicate this result in Fig.~2(a), in which $\phi_0$ is always $24.12^{\circ}$.

By taking the dimensionless force at $s=0$, i.e., $P_0R^2/B=1/\sin\phi_1=1.22$, into the curvature equation for the detached strip, the detached length can be numerically found to be $L^\prime/R=2.088$. 
This ratio also reflects the minimum length permitted in the continuous-contact deformation, i..e, $2L^\prime$.
However, the continuous-contact deformation will be terminated when two free ends of the strip meet at $(X/R,Y/R)=(0,0)$.
Accordingly, we can derive the maximum length for the continuous contact situation by $2L^{'}+R(2\pi-2\phi_1)=6.091R$.
Therefore, the supported strip length for this continuous-contact region is bounded by
 \begin{equation}
 L/R \in [4.176 ,6.091]; \, \text{continuous-contact region}.
 \end{equation}
 We want to emphasize that  the corresponding dimensionless force $P_0R^2/B$ in this continuous-contact region, somewhat  counter-intuitively, remains constant, as the numerical results show in Fig.~2 (b).\\

\noindent {\it Self-contact.}\textemdash   Last but not least, we consider the strip  length in the self-contact region:
\begin{equation}
L/R >6.091;\, \text{self-contact region}.
\end{equation}
Now, the  strip self-contacts, as illustrated  in Figs.~1(d) and 1(h) from simulation and experiment, respectively. 
Here, one free end of the strip makes a point-contact with the front side of the tube.
At this stage, the point-contact is accompanied with internal point forces, as denoted by $\vec{P}_1$ and $\vec{P}_4$ shown in Fig.~1(d).
By Newton's third law, we have $\vec{P}_1=-\vec{P}_4$. As a result, the interaction among different segmental parts of the strip leads to a nonlocal effect.
Moreover, there still exists a finite continuous-contact region, marked by the green colored curve in Fig.~1(d), in association with the pressure distribution $\vec{K}$.

Interestingly, only two external point forces are upon the whole strip. 
One is the point-contact force at $s=0$, denoted as $\vec{P}_0$, and the other one is the pressure $\vec{K}$ from the tube.
Since the null of friction is still valid in our system, the direction of the point contact force at $s=0$ is  vertical.
Consequently, in order to maintain the static equilibrium, the pressure distribution $\vec{K}$ needs to be a symmetric  distribution with respect to Y-axis.
This implies that the positions of $C_1$ and $C_2$ form mirror symmetry with respect to the Y-axis.  The detailed derivation can be found in  the Supplementary Materials.

In Fig.~2 (b), we also calculate the corresponding dimensionless point-contact force at $s=0$. 
Since the emergence of internal force is close to $s=0$ in our case, the dimensionless point-contact force $P_0R^2/B$ can be expectedly to be larger than that in the continuous-contact region. 
At $L/R=2\pi$, we find $P_0R^2/B=2.76$.
Consequently, the pushing force $\vec{P}_1$ causes the front strip downward, resulting in $\phi_0$  decreasing  with respect to $L/R$, as shown in Fig.~2(a).\\

\vspace{0.2in}
\begin{table}[t]
  \begin{tabular} {| p {2.0 cm} | p {1.3 cm} | p {1.3 cm} |   p {1.3 cm} |    p {1.8 cm} |}
    \hline
    Length Ratio $L/R$  & PVC $0.5$mm  & PET $0.2$mm  & PVC $0.1$mm  & Theoretical Values, $\phi_0$\\
    \hline
    2.4 & $43.9^{\circ}$ & $42.8^{\circ}$ & $42.3^{\circ}$ & $42.37^{\circ}$\\
    2.6 & $33.1^{\circ}$ & $34.2^{\circ}$ & $34.1^{\circ}$ & $33.55^{\circ}$\\
    2.8 & $26.7^{\circ}$& $28.6^{\circ}$ & $28.1^{\circ}$ & $26.78^{\circ}$\\
    3.03 & $20.8^{\circ}$& $20.2^{\circ}$ & $20.7^{\circ}$ & $20.5^{\circ}$\\
    3.2 & $21.2^{\circ}$& $21.3^{\circ}$ & $21.3^{\circ}$ & $21.38^{\circ}$\\
    3.7 & $23.3^{\circ}$& $23.2^{\circ}$ & $24^{\circ}$ & $23.38^{\circ}$\\
    4.176 & $23.4^{\circ}$& $23.6^{\circ}$  & $24.3^{\circ}$ & $24.12^{\circ}$\\
    4.5 & $24.2^{\circ}$ & $24.3^{\circ}$  & $24.3^{\circ}$ & $24.12^{\circ}$\\
    5.1 & $23.8^{\circ}$ & $24.2^{\circ}$ & $24.1^{\circ}$& $24.12^{\circ}$\\  
    6.28 & $17.5^{\circ}$& $17.5^{\circ}$ & $17.9^{\circ}$ & $15.76^{\circ}$\\
    \hline
         \end{tabular}
     \caption{Experimental measurements on the tangential angle $\phi_0$ from  different stripe length ratio $L/R$. Here, we have  three sets in material parameters: $0.5$mm thickness of PVC (PVC1), $0.2$ mm thickness of PET (PET), and $0.2$mm thickness of PVC (PVC2). Theoretical values are also listed for the comparison.}
\end{table}


\noindent {\it Experimental verification.}\textemdash We design and fabricate a series of different stripe lengths $L/R$, from $2.4$ to $6.28$, see Table 1 for more details.
In experiments, we prepare two different elastic materials: polyvinyl chloride (PVC) and  polyethylene terephthalate (PET), but also with different thickness  in order to verify our theoretical findings.
Three sets of material parameters are performed:  $0.5$mm thickness of PVC (PVC1), $0.2$ mm thickness of PET (PET), and $0.1$mm thickness of PVC (PVC2).
All the samples are $2$ cm wide.
The elastic strips  are also prepared in initially flat condition, i.e.,  in the absence of tube confinement, to avoid any plastic deformation.
The tube used is acrylic (polymethylmethacrylate, PMMA). It  an inner radius of $3$ cm.

With the tangential angle at $s=0$, denoted as $\phi_0$, the measured tangential angles from three sets of PVC1, PET, and PVC2, are listed in Table 1, along with the comparison to the theoretical values. 
All the obtained data are also plotted in Fig.~2(a), as well as the selected pictures shown in Fig.~1(e)-(h).
They show good agreements with our simulation curves for two point-contact, three point-contact, and continuous-contact regions.
When the length of strip satisfies $4.176\leq L/R\leq 6.091$, the measured $\phi_0$ confirms the theoretical value $24.12^{\circ}$ according to prediction from the continuous contact case.\\


\noindent {\it Conclusion.}\textemdash 
In addition to the universal law of coiling for a long stripe, we find theoretically and experimentally that a short elastic strip contacting within a tube, with the length ratio  $2 < L/R \le 2\pi$,  also follows  universal behavior.
Four different types of deformation: two point-contact, three point-contact, continuous-contact, and self-contact, are identified in theory and verified in experiments.
Theoretically, the boundaries between two adjunct regions of deformation are characterized by elastic Kirchhoff's equations; while experimentally three sets of material parameters, with a series of different lengths, are investigated,  resulting in good agreement with our theoretical analysis.
Our results show the existence of a universal law even for a short strip, irrespective of elastic properties, strip lengths, and tube radii.
The results in this work can be readily applied to many practical applications, ranging, e.g., from flexible electronic devices, medical fibre imaging, to DNA packaging.

\section*{Acknowledgement}
The authors are indebted to Prof. Ole Stuernagle for useful discussions.
This work is partially supported by the Ministry of Science and Technology of Taiwan (Nos. 110-2112-M-259-005,  111-2112-M-259-011, 110-2627-M-008-001, and 110-2123-M-007-002), the International Technology Center Indo-Pacific (ITC IPAC) and Army Research Office, under Contract No. FA5209-21-P-0158, and the Collaborative research program of the Institute for Cosmic Ray Research (ICRR), the University of Tokyo.


 \section{Two point-contact region}
 
 \begin{figure*}[ht]
\centering
\includegraphics[width=1.0\textwidth]{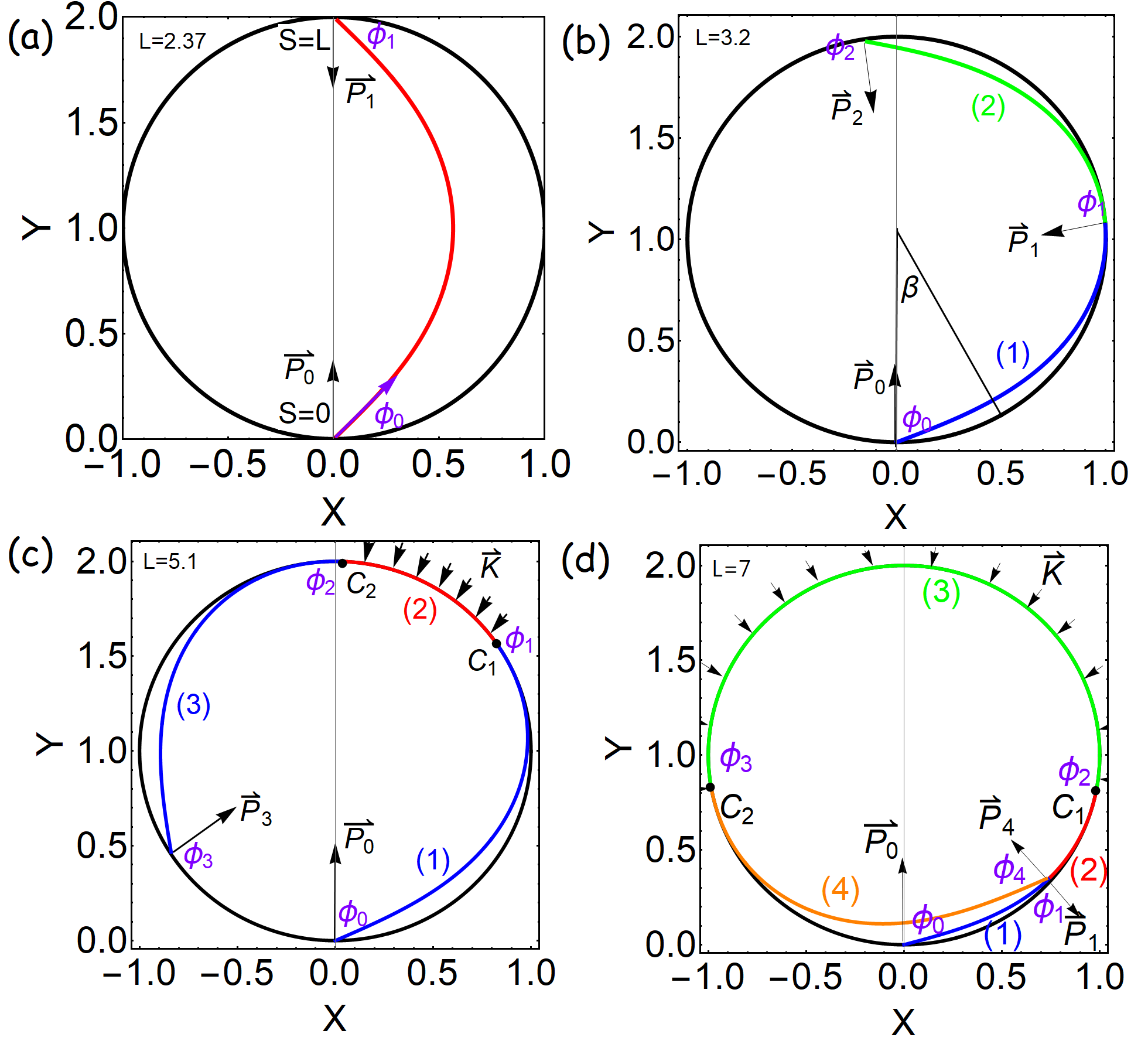}
\caption{(a) Deformation  of two point-contact by numerical calculation with $L/R=2.37$. The external forces normal to the tube wall at $s=0$ and $s=L$ are  denoted as $\vec{P}_0$ and $\vec{P}_1$, respectively. Here, $\phi_0$ and $\phi_1$ are the tangential angles with respect to the X-axis at $s=0$ and $s=L$. (b) Deformation of three point-contact by numerical calculation with $L/R=3.2$. There are three external point-contact forces upon the strip, denoted as $\vec{P}_0$, $\vec{P}_1$, and $\vec{P}_0$. The corresponding tangential angles of the strip at $s=0$, $s=\frac{L}{2}$, and $s=L$, are $\phi_0$, $\phi_1$, and $\phi_2$, respectively. Here, we also employ another coordinate system $\beta$,  with respect to the $-$Y-axis, as an auxiliary coordinate to describe the geometric relation. (c) Deformation of continuous-contact by $L/R=5.1$. There exists a finite continuous-contact region marked by a red curve, in association with the continuous contact pressure $\vec{K}$ marked by black arrows.  Here, $C_1$ and $C_2$ correspond to the boundary of the continuous contact region. $\phi_0$, $\phi_1$, $\phi_2$, and $\phi_3$ are the corresponding tangential angles at $s=0$, $C_1$, $C_2$, and $s=L$, respectively. (d) Deformation of self-contact by $L=7R$. There still exists a continuous-contact region marked by a green curve, in association with pressure $\vec{K}$. Here, the external forces are from $\vec{P}_0$ at $s=0$ and pressure $\vec{K}$ occurred at the continuous-contact region.  $\vec{P_1}$ and $\vec{P}_4$ as internal forces. The regions, (1), (2), and (4) are detached from the tube, marked by blue, red, and orange curves. }
\label{fig-SM}
\end{figure*}

 Figure~\ref{fig-SM}(a) gives an example of the two point-contact scenario. Here, we have $L/R=2.37$ in numerical calculation.
Here, the origin of coordinate (X,Y)  is referred to  the bottom of the tube, where is also the onset of the strip, i.e., $s=0$.
 Related force analysis diagram is also illustrated in Fig.~\ref{fig-SM} (a). There are two external point-contact forces,  denoted as $\vec{P}_0$ and $\vec{P}_1$, upon two ends of the strip.
  
 From  the force balance, we obtain $\vec{P}_0=-\vec{P}_1$.  And due to free of friction, we have $\vec{P}_0=P_0\hat{y}$.
    We denote $\phi_0$ and $\phi_1$ as the corresponding tangential angles of the strip with respect to X-axis at $s=0$ and $s=L$, respectively.
 In our coordinate, the unit tangential vector has the form:  $\hat{t}=\frac{dx}{ds}\hat{x}+\frac{dy}{ds}\hat{y}=\cos\phi(s)\hat{x}+\sin\phi(s)\hat{y}$.
As we only have two-point contacts, from Eq.~(1) in the main text, we read $\vec{F}_s=0$. Therefore the stress  force at each cross section of the strip is constant, i.e., $\vec{F}=-\vec{P}_0=-P_0\hat{y}$.

Resorting to the moment balance of Eq.~(1), we obtain a curvature equation,
\begin{equation}\label{1}
\frac{d\phi(s)}{ds}=\phi(s)_s=\sqrt{\frac{2P_0}{B}(\sin\phi-\sin\phi_0)}.
\end{equation}
Here we already apply  a zero moment at $s=0$, corresponding to $\phi_s(0)=0$.
Since the length of the strip projected  to Y-axis is always $R$ and the strip length $L$ is conserved, we thus have two geometric constraints:
\begin{eqnarray}\label{2}
R&=\int_{\phi_0}^{\frac{\pi}{2}}\frac{\sin\phi}{\sqrt{\frac{2P_0}{B}(\sin\phi-\sin\phi_0)}}d\phi,\\
\frac{L}{2}&=\int_{\phi_0}^{\frac{\pi}{2}}\frac{d\phi}{\sqrt{\frac{2P_0}{B}(\sin\phi-\sin\phi_0)}}d\phi,
\end{eqnarray}
in which  two unknowns $\phi_0$ and $P_0$ are involved. With above equations, one can eliminate $P_0$ and reach
\begin{equation}\label{lengthtwo}
\frac{2R}{L}=\frac{\int_{\phi_0}^{\frac{\pi}{2}}\frac{\sin\phi}{\sqrt{\sin\phi-\sin\phi_0}}d\phi}{\int_{\phi_0}^{\frac{\pi}{2}}\frac{d\phi}{\sqrt{\sin\phi-\sin\phi_0}}}.
\end{equation}
Therefore we determine $\phi_0$ with a given value of $L/R$.

Now we define a dimensionless force by $\frac{P_0R^2}{B^2}$, which has the form: 
\begin{equation}\label{pointforce}
\frac{P_0R^2}{B}=\frac{1}{2}[\int_{\phi_0}^{\frac{\pi}{2}}\frac{\sin\phi}{\sqrt{\sin\phi-\sin\phi_0}}d\phi]^2,
\end{equation}
as a  function of $\phi_0$.

We remark that in this two point-contact region,  a crucial geometric constraint is that  the position of the middle part of the strip projected to x-axis is bound by
\begin{eqnarray}\label{condition}
0\leqslant&x\leqslant R  \quad \Rightarrow\quad  1-\frac{1}{4}[\int_{\phi_0}^{\frac{\pi}{2}}\frac{\sin\phi}{\sqrt{\sin\phi-\sin\phi_0}}d\phi]^2 &\leq \sin\phi_0\leq 1.
\end{eqnarray}
Here,  we already apply  $P_0$ from Eq.~(\ref{pointforce}) and $x=\int_{\phi_0}^{\frac{\pi}{2}}\frac{\cos\phi}{\sqrt{\frac{2P_0}{B}(\sin\phi-\sin\phi_0)}}d\phi=\sqrt{\frac{2B}{P_0}[1-\sin\phi_0]}$. 
By solving above condition, we then obtain
\begin{equation}
0.358 \leq \phi_0<\frac{\pi}{2}.
\end{equation}
Next, combined with Eq.~(\ref{lengthtwo}), the strip length in this two point-contact region is bounded by
\begin{equation}
L/R \in [2, 3.033),
\end{equation}
which is valid for any elastic materials.

Once  we determine $\phi_0$ and $P_0$, the shape of elastic strip can be numerically calculated through
\begin{eqnarray}
x(\phi^{'})&=&\int_{\phi_0}^{\phi^{'}}\frac{\cos\phi}{\sqrt{\frac{2P_0}{B}(\sin\phi-\sin\phi_0)}}d\phi
=\sqrt{\frac{2B}{P_0}[\sin\phi^{'}-\sin\phi_0]},\\
y(\phi^{'})&=&\int_{\phi_0}^{\phi^{'}}\frac{\sin\phi}{\sqrt{\frac{2P_0}{B}(\sin\phi-\sin\phi_0)}}d\phi,
\end{eqnarray}
where  $\phi^{'}=[\phi_0,\pi-\phi_0]$. An example of two point-contact deformation for  $L/R=2.37$ is illustrated in Figure~\ref{fig-SM}(a).

 \section{Three point-contact region}
When $L/R > 3.033$, the strip encounters three external forces, denoted as $\vec{P}_0$, $\vec{P}_1$, and $\vec{P}_2$. 
 The corresponding tangential angles of the strip with respect to X-axis are denoted as $\phi_0$, $\phi_1$, and $\phi_2$, respectively.
 We note that in this stage the curvature of the strip at the middle part is larger than $1/R$, corresponding to a point contact.
On the other hand, from the force balance, it leads to $\vec{P}_0+\vec{P}_1+\vec{P}_2=0$. 
Now, we introduce another angle coordinate $\beta$, with respect to $-$Y axis, as shown in Fig.~\ref{fig-SM}(b), as an auxiliary coordinate.
Thus, the coordinates of these contact-points can be written as
\begin{equation}\label{position}
\vec{R}(\beta)=R\sin\beta\hat{x}+(R-R\cos\beta)\hat{y},
\end{equation}
with $\beta=[\beta_0,\beta_1,\beta_2]$.
With the auxiliary coordinate, we have $\beta_1=\phi_1$, but $\phi_0\neq \beta_0$ and $\phi_2\neq \beta_2$.
As a result, the corresponding point-contact forces are
\begin{eqnarray}
\vec{P}_0&=&P_0\hat{y},\\
\vec{P}_1&=&P_1(-\sin\beta_1\hat{x}+\cos\beta_1\hat{y}),\\
\vec{P}_2&=&P_0(-\sin\beta_2\hat{x}+\cos\beta_2\hat{y}).
\end{eqnarray}

With mirror symmetry, we have $\vert \vec{P}_0\vert=\vert \vec{P}_2\vert=P_0$. Then, by force balance, one obtain $\vec{P}_0+\vec{P}_1+\vec{P}_2=0$, or equivalently, 
\begin{eqnarray}
&&\hat{x}: -P_1\sin\beta_1=P_0\sin\beta_2\\
&&\hat{y}:P_0+P_1\cos\beta_1+P_0\cos\beta_2=0.
\end{eqnarray}
Then, we find
\begin{eqnarray}\label{open1}
\tan\beta_1&=\frac{\sin\beta_2}{1+\cos\beta_2},\\
P_1&=-\frac{P_0\sin\beta_2}{\sin\beta_1}.
\end{eqnarray}
Once we determine $\beta_1$ ($\phi_1$) and $P_0$,  $\beta_2$ and $P_1$ can be  obtained.
 
 Further more, we divide the detached strips into two regions, denoted as $(1)$ and $(2)$, marked by blue and green curves in Fig.~\ref{fig-SM}(b).
From the moment equation, a curvature equation for the region (1) can be obtained, i.e., 
 \begin{eqnarray}\label{curve1}
\phi^{(1)}_s=&\sqrt{\frac{2P_0}{B}(\sin\phi-\sin\phi_0)}.
 \end{eqnarray}
Here we also apply  a zero moment at $s=0$, corresponding to $\phi_s(0)=0$.
Complemented by the geometry constraint for the middle part of the strip, we  have the following conditions,
\begin{eqnarray}\label{position1}
R\sin\phi_1&=&\int_{\phi_0}^{\phi_1}\frac{\cos\phi d\phi}{\sqrt{\frac{2P_0}{B}[\sin\phi-\sin\phi_0]}},\\
R-R\cos\phi_1&=&\int_{\phi_0}^{\phi_1}\frac{\sin\phi d\phi}{\sqrt{\frac{2P_0}{B}[\sin\phi-\sin\phi_0]}}.
\end{eqnarray} 
Again, by eliminating the  unknown $P_0$, we have
\begin{equation}\label{comb}
\frac{1-\cos\phi_1}{\sin\phi_1}=\frac{\int_{\phi_0}^{\phi_1}\frac{\sin\phi d\phi}{\sqrt{\sin\phi-\sin\phi_0}}}{\int_{\phi_0}^{\phi_1}\frac{\cos\phi d\phi}{\sqrt{\sin\phi-\sin\phi_0}}},
\end{equation}
with  two unknowns involved, i.e., $\phi_0$ and $\phi_1$. For this segment of the strip, the corresponding length is
\begin{equation}\label{length}
\frac{L}{2}=\int_{\phi_0}^{\phi_1}\frac{d\phi}{\sqrt{\frac{2P_0}{B}[\sin\phi-\sin\phi_0]}}.
\end{equation}
With Eqs.~(\ref{position1}) and (\ref{length}), we have
\begin{eqnarray}\label{length1}
\int_{\phi_0}^{\phi_1}\frac{d\phi}{\sqrt{\sin\phi-\sin\phi_0}}=\frac{L}{R}\frac{\sqrt{\sin\phi_1-\sin\phi_0}}{\sin\phi_1}.
\end{eqnarray}
Then, by considering a set of Eqs.~(\ref{comb}) and (\ref{length1}):
 \begin{eqnarray}\label{three}
 \frac{1-\cos\phi_1}{\sin\phi_1}&=\frac{\int_{\phi_0}^{\phi_1}\frac{\sin\phi d\phi}{\sqrt{\sin\phi-\sin\phi_0}}}{\int_{\phi_0}^{\phi_1}\frac{\cos\phi d\phi}{\sqrt{\sin\phi-\sin\phi_0}}},\\
 \int_{\phi_0}^{\phi_1}\frac{d\phi}{\sqrt{\sin\phi-\sin\phi_0}}&=\frac{L}{R}\frac{\sqrt{\sin\phi_1-\sin\phi_0}}{\sin\phi_1},
 \end{eqnarray}
one can simultaneously solve $\phi_0$ and $\phi_1$ ($\beta_1$) when $L/R$ is given.

The corresponding dimensionless force $\frac{P_0R^2}{B}$ for the three point-contact case has the form:
 \begin{equation}\label{p0}
 \frac{P_0R^2}{B}=\frac{2(\sin\phi_1-\sin\phi_0)}{\sin^2\phi_1},
 \end{equation}
 with an implicit dependence of $L/R$.
 
 Regarding the other region (2), i.e., $s=[\frac{L}{2},L]$,  we have $\vec{F}=\vec{P}_2=P_0(-\sin\beta_2\hat{x}+\cos\beta_2\hat{y})$, where $\beta_2$ can be calculated by using Eq.~(\ref{open1}). Again, by using moment equation, the curvature equation for the region (2) is
\begin{widetext}
\begin{eqnarray} \label{curve2}
\phi^{(2)}_s=\sqrt{\frac{2}{B}[P_0\sin(\beta_2-\phi)+\frac{2B(\sin\phi_1-\sin\phi_0)^2}{R^2\sin^2\phi_1}-P_0\sin(\beta_2-\phi_1)]},
\end{eqnarray}
\end{widetext}
where we apply the condition of a continuous curvature at $\phi_1$, i.e.,  between the region (1) and the region (2).
The curvature at $s=L/2$ is
\begin{equation}\label{universal}
\phi_s(\frac{L}{2})=\frac{2(\sin\phi_1-\sin\phi_0)}{R\sin\phi_1}.
\end{equation}
 For this detached region (2), the corresponding length is defined by 
\begin{widetext} 
\begin{equation}
\frac{L}{2}=\int_{\phi_1}^{\phi_2}\frac{d\phi}{\sqrt{\frac{2}{B}[P_0\sin(\beta_2-\phi)+\frac{2B(\sin\phi_1-\sin\phi_0)^2}{R^2\sin^2\phi_1}-P_0\sin(\beta_2-\phi_1)]}},
\end{equation}
\end{widetext}
which can be used to determine $\phi_2$, i.e., tangential angle at $s=L$.

The three-point contact region will be  terminated when the curvature at the middle plate meets $\phi_s(\frac{L}{2})=\frac{2(\sin\phi_1-\sin\phi_0)}{R\sin\phi_1}=\frac{1}{R}$, resulting in the continuous-contact region.
Consequently,  the strip length for three-point contact situation is bounded by
\begin{equation}
L/R \in[3.033R,4.176),
\end{equation}
which is again valid for any elastic materials.
In Fig.~\ref{fig-SM}(b), we illustrate such a  three point-contact deformation from the numerical calculation with   $L/R=3.2$.

 \section{Continuous-contact region}

When $L/R > 4.176$, we enter the continuous-contact region, as  illustrated in  Fig.~\ref{fig-SM}(c).
Here,  the boundaries of the continuous-contact are defined as  $C_1$ and $C_2$, with its tangential angles as $\phi_1$ and $\phi_2$, respectively.
The corresponding force analysis diagram is also shown in Fig.~\ref{fig-SM}(c).
Now, there are two point-contact forces $\vec{P}_0$ and $\vec{P}_3$ and a continuous-contact pressure $\vec{K}$. 
Here, $\vec{K}$ is pressure defined as force per unit length.
The tangential angles of the strip at $\vec{P}_0$ and $\vec{P}_3$ with respect to X-axis  are denoted as  $\phi_0$ and $\phi_3$, respectively.

By considering mirror symmetry for the strip, we have  $\vert \vec{P}_0\vert=\vert \vec{P}_3\vert=P_0$ and the detached shapes, denoted as (1) and (3), are expected to be identical.
It is noted that within this continuous-contact region, its curvature remains $1/R$.
Bu a properly chosen coordinate, we also have $\phi_1=\beta_1$ and $\phi_2=\beta_2$.
Now, for the region (1), the stress force is constant, i.e., $\vec{F}=-\vec{P}_0=-P_0\hat{y}$. with the help of  moment balance, we read
\begin{equation}
\phi_s^{(1)}=\sqrt{\frac{2P_0}{B}(\sin\phi-\phi_0)},
\end{equation}
where a zero moment at $s=0$, corresponding to $\phi_s(0)=0$ is used.
As the curvature at $C_1$ is $\frac{1}{R}$, one can have
\begin{equation}\label{3}
\frac{1}{R}=\sqrt{\frac{2P_0}{B}(\sin\phi_1-\phi_0)}.
\end{equation}
Since $C_1$ is confined by the tube, we therefore have the geometric constraints:
\begin{widetext}
\begin{eqnarray}\label{4}
R\sin\phi_1&=&\int_{\phi_0}^{\phi_1}\frac{\cos\phi}{\sqrt{\frac{2P_0}{B}(\sin\phi-\sin\phi_0)}}d\phi =\sqrt{\frac{2B}{P_0}(\sin\phi_1-\sin\phi_0)},\\
R-R\cos\phi_1&=&\int_{\phi_0}^{\phi_1}\frac{\sin\phi}{\sqrt{\frac{2P_0}{B}(\sin\phi-\sin\phi_0)}}d\phi.
\end{eqnarray}
\end{widetext}
By eliminating $P_0$, we reach the following equation
\begin{equation}\label{univ1}
\frac{\sin\phi_1}{1-\cos\phi_1}=\frac{\int_{\phi_0}^{\phi_1}\frac{\cos\phi}{\sqrt{(\sin\phi-\sin\phi_0)}}d\phi}{\int_{\phi_0}^{\phi_1}\frac{\sin\phi}{\sqrt{(\sin\phi-\sin\phi_0)}}d\phi}.
\end{equation}
By taking Eq.~(\ref{3}) into the first  term of Eq.~(\ref{4}), we obtain
\begin{equation}\label{univ2}
2\sin\phi_0=\sin\phi_1.
\end{equation}
Consequently, we have a set of  Eqs.~(\ref{univ1}) and (\ref{univ2}),
\begin{eqnarray}\label{univ3}
2\sin\phi_0&=&\sin\phi_1,\\
\frac{\sin\phi_1}{1-\cos\phi_1}&=&\frac{\int_{\phi_0}^{\phi_1}\frac{\cos\phi}{\sqrt{(\sin\phi-\sin\phi_0)}}d\phi}{\int_{\phi_0}^{\phi_1}\frac{\sin\phi}{\sqrt{(\sin\phi-\sin\phi_0)}}d\phi},
\end{eqnarray}
where two unknowns, i.e., $\phi_0$ and $\phi_1$ are involved, which can be sufficiently determined, independent from  the elastic bending moment $B$ and tube radius $R$.
Numerically,  we obtain
\begin{eqnarray}
\phi_0&=0.421=24.1^{\circ},\\
\phi_1&=2.185=125.2^{\circ}.
\end{eqnarray}
The results reveal that the tangential angle of the free end is $\phi_0=0.421=24.12^{\circ}$; while the opening angle is  $\phi_1=\beta_1=2.185=125.2^{\circ}$, valid for any elastic material and tube size, referring to universal law of coiling.


The corresponding dimensionless force $P_0R^2/B$ in this continuous contact region  is
\begin{equation}\label{univ4}
\frac{P_0R^2}{B}=\frac{1}{\sin\phi_1}=1.22.
\end{equation}
 
Moreover, we formulate the detached length of the region (1), defined as $L^{'}$, 
\begin{widetext}
\begin{eqnarray}
 \frac{L^{'}}{R}=\frac{1}{R} \int_{\phi_0}^{\phi_1}\frac{d\phi}{\sqrt{\frac{2P_0}{B}(\sin\phi-\sin\phi_0)}} =\sqrt{\frac{\sin\phi_1}{2}}\int_{\phi_0}^{\phi_1}\frac{d\phi}{\sqrt{\sin\phi-\sin\phi_0}} =2.088.
\end{eqnarray}
\end{widetext}
Here, we already apply Eq.~(\ref{univ4}) with $\phi_0=0.421$ and  $\phi_1=2.185$.
This result defines the minimum length for the emergence of the continuous contact, i.e., $2L^{'}$.
On the other hand, the maximum length before the self-contact situation occurs with
\begin{equation}
2\frac{L^{'}}{R}+(2\pi-2\beta_1).
\end{equation}
Here, $2L^{'}/R$ denotes the total length of detached region; while $(2\pi-2\beta_1)$ denotes the length of the continuous-contact region.
As a result the strip length for this continuous-contact is bounded by
 \begin{equation}
 L \in [4.176 R,6.091R],
 \end{equation}
which is valid for any elastic materials.

 For the region (3),  due to mirror symmetry of the strip, $\beta_2$ (here $\beta_2=\phi_2$) can be found by
 \begin{equation}
 \beta_2=\frac{L-2L^{'}}{R}+\beta_1.
 \end{equation}
By geometry, we have
 \begin{equation}
 \beta_3=\beta_1+\beta_2.
 \end{equation}
We note that the tangential angle at $s=L$ end, i.e.,$\phi_3$, is not the same as $\beta_3$. 
By using the detached length condition, we can have  $\phi_3$ by
 \begin{equation}
\frac{L^{'}}{R}=\frac{1}{R} \int_{\phi_2}^{\phi_3}\frac{d\phi}{\sqrt{\frac{2P_0}{B}(\sin[\phi_3-\beta_3]-\sin[\phi-\beta_3])}}=2.088.
 \end{equation}
The corresponding curvature equation for this detached region (3) has the form:
 \begin{equation}
 \phi_s^{(3)}=\sqrt{\frac{2P_0}{B}[\sin(\phi_3-\beta_3)-\sin(\phi-\beta_3)]}.
 \end{equation}
The stress force at this detached region is $\vec{F}^{(3)}=q_0(-\sin\beta_3\hat{x}+\cos\beta_3\hat{y})$.
An example of this continuous contact region is illustrated in   
   Fig. ~\ref{fig-SM}(c) with numerical calculation for $L/R=5.1$.

  \section{Self-contact region}
The deformation of self-contact occurs at
\begin{equation}
L/R > 6.091.
\end{equation}
 Due to the interaction among different segments of the strip emerges, the self-contact problem is nonlocal and highly nonlinear.
As illustrated in  Fig. ~\ref{fig-SM}(d), we divide the strip into four segments, denoted as (1), (2), (3), and (4), with the corresponding length $L_1$, $L_2$, $L_3$, and $L_4$, respectively.
First of all, we have  $L=L_1+L_2+L_3+L_4$.
Here, only the region (3) is subject to the continuous contact.
We also denote  $\phi_0$, $\phi_1$, $\phi_2$, $\phi_3$, and $\phi_4$ are the tangential angles with respect to X-axis at $s=0$ and the interface between regions (1)-(2), $C_1$, $C_2$, and $s=L$.
   Here $C_1$ and $C_2$ are the boundaries of the continuous contact.
   
With the help of force analysis diagram in Fig.~\ref{fig-SM}(d), the external forces are $\vec{P}_0$ and pressure $\vec{K}$; while $\vec{P_1}$ and $\vec{P}_4$ belong to internal forces.
The continuous contact region is associated with the pressure $\vec{K}$ by the tube with a constant in magnitude.
On the other hand, since the direction of $\vec{P}_0$ is always toward Y-axis direction, in order to maintain the force balance, we can conclude that the pressure $\vec{K}$ has to be symmetry with respect to Y-axis.
Accordingly, a geometric relation for $\beta_2$ and $\beta_3$ can be constructed, i.e., 
\begin{eqnarray}
 \pi-\beta_2=\beta_3-\pi \quad \Rightarrow \quad \beta_2+\beta_3&=2\pi.
\end{eqnarray}

In the region (1), i.e., $s\in[0, L_1]$, the stress force is $\vec{F}^{(1)}=-\vec{P}_0=-P_0\hat{y}$ and the corresponding curvature equation is $\phi_s^{(1)}=\sqrt{\frac{2P_0}{B}[\sin\phi-\sin\phi_0]}$.

Then, in the region (2), i.e., $s\in[L_1,L_2+L_1]$,  the stress force is $\vec{F}^{(2)}=-\vec{P}_0-\vec{P}_1=-P_1\sin\phi_1\hat{x}+[-P_0+P_1\cos\phi_1]\hat{y}$. 
The corresponding curvature equation is   $\phi_s^{(2)}=\sqrt{\frac{2}{B}[P_0\sin\phi-P_1\sin(\phi-\phi_1)-P_0\sin\phi_0]}$, where we apply the boundary conditions at $s=L_1$ and $s=L_2$.
 
Finally, in the region (4), the stress force is $\vec{F}^{(4)}=P_1(-\sin\phi_1\hat{x}+\cos\phi_1\hat{y})$. By Newton's third law, we have $\vec{P}_1=-\vec{P}_4$.
The curvature equation is
$\phi_s^{(4)}=\sqrt{\frac{2P_1}{B}[\sin(\phi_4-\phi_1)-\sin(\phi-\phi_1)}$, where  we also  apply  a zero moment at $s=L$, i.e., $\phi_s(L)=0$. Now, we have  six unknowns in this self-contact problem. They are 
 \begin{equation}
 \phi_0,\phi_1,\beta_2,\phi_4,P_0,P_1.
 \end{equation}
To solve, we need six equations. First, the position at $s=L_1$ meets that at $s=L$. Then, we have
\begin{widetext}
  \begin{eqnarray}
 \int_{\phi_0}^{\phi_1}\frac{\cos\phi d\phi}{\sqrt{\frac{2P_0}{B}[\sin\phi-\sin\phi_0]}} &=& R\sin(\beta_3)+\int_{\beta_3}^{\phi_4}\frac{\cos\phi d\phi}{\sqrt{\frac{2P_1}{B}[\sin(\phi_4-\phi_1)-\sin(\phi-\phi_1)]}},\\
\int_{\phi_0}^{\phi_1}\frac{\sin\phi d\phi}{\sqrt{\frac{2P_0}{B}[\sin\phi-\sin\phi_0]}} &=& R-R\cos(\beta_3)+\int_{\beta_3}^{\phi_4}\frac{\sin\phi d\phi}{\sqrt{\frac{2P_1}{B}[\sin(\phi_4-\phi_1)-\sin(\phi-\phi_1)]}}.
  \end{eqnarray}
 \end{widetext}
In addition, we also have a position condition at $s=L_1+L_2$, i.e., 
\begin{widetext}
  \begin{eqnarray}
R\sin\beta_2 &=& \int_{\phi_0}^{\phi_1}\frac{\cos\phi d\phi}{\sqrt{\frac{2P_0}{B}[\sin\phi-\sin\phi_0]}}+\int_{\phi_1}^{\beta_2}\frac{\cos\phi d\phi}{\sqrt{\frac{2}{B}[P_0(\sin\phi-\sin\phi_0)-P_1\sin(\phi-\phi_1)]}},\\
  R-R\cos\beta_2 &=& \int_{\phi_0}^{\phi_1}\frac{\sin\phi d\phi}{\sqrt{\frac{2P_0}{B}[\sin\phi-\sin\phi_0]}}+\int_{\phi_1}^{\beta_2}\frac{\sin\phi d\phi}{\sqrt{\frac{2}{B}[P_0(\sin\phi-\sin\phi_0)-P_1\sin(\phi-\phi_1)]}}.
  \end{eqnarray}
   \end{widetext}
  We also have the total strip length $L$ as
\begin{widetext}
  \begin{eqnarray}
  L&=&\int_{\phi_0}^{\phi_1}\frac{ d\phi}{\sqrt{\frac{2P_0}{B}[\sin\phi-\sin\phi_0]}}+\int_{\phi_1}^{\beta_2}\frac{ d\phi}{\sqrt{\frac{2}{B}[P_0(\sin\phi-\sin\phi_0)-P_1\sin(\phi-\phi_1)]}}\\
  &+&R(\beta_3-\beta_2)+\int_{\beta_3}^{\phi_4}\frac{ d\phi}{\sqrt{\frac{2P_1}{B}[\sin(\phi_4-\phi_1)-\sin(\phi-\phi_1)]}}.
  \end{eqnarray}
  \end{widetext}
  With the  conservation relation, one can obtain
 \begin{equation}
 \frac{B}{2R^2}-P_0\sin\phi_2+P_1\sin(\phi_2-\phi_1)=-P_0\sin\phi_0.
 \end{equation}
Finally, we summarize a set of six equations for these six unknowns:
 \begin{widetext}
 \begin{eqnarray}
 &&\int_{\phi_0}^{\phi_1}\frac{\cos\phi d\phi}{\sqrt{\frac{2P_0}{B}[\sin\phi-\sin\phi_0]}}= R\sin(\beta_3)+\int_{\beta_3}^{\phi_4}\frac{\cos\phi d\phi}{\sqrt{\frac{2P_1}{B}[\sin(\phi_4-\phi_1)-\sin(\phi-\phi_1)]}},\\
&& \int_{\phi_0}^{\phi_1}\frac{\sin\phi d\phi}{\sqrt{\frac{2P_0}{B}[\sin\phi-\sin\phi_0]}}= R-R\cos(\beta_3)+\int_{\beta_3}^{\phi_4}\frac{\sin\phi d\phi}{\sqrt{\frac{2P_1}{B}[\sin(\phi_4-\phi_1)-\sin(\phi-\phi_1)]}},\\
R\sin\beta_2&=&\int_{\phi_0}^{\phi_1}\frac{\cos\phi d\phi}{\sqrt{\frac{2P_0}{B}[\sin\phi-\sin\phi_0]}}+\int_{\phi_1}^{\beta_2}\frac{\cos\phi d\phi}{\sqrt{\frac{2}{B}[P_0(\sin\phi-\sin\phi_0)-P_1\sin(\phi-\phi_1)]}},\\
 R-R\cos\beta_2&=&\int_{\phi_0}^{\phi_1}\frac{\sin\phi d\phi}{\sqrt{\frac{2P_0}{B}[\sin\phi-\sin\phi_0]}}+\int_{\phi_1}^{\beta_2}\frac{\sin\phi d\phi}{\sqrt{\frac{2}{B}[P_0(\sin\phi-\sin\phi_0)-P_1\sin(\phi-\phi_1)]}},\\
   L&=&\int_{\phi_0}^{\phi_1}\frac{ d\phi}{\sqrt{\frac{2P_0}{B}[\sin\phi-\sin\phi_0]}}+\int_{\phi_1}^{\beta_2}\frac{ d\phi}{\sqrt{\frac{2}{B}[P_0(\sin\phi-\sin\phi_0)-P_1\sin(\phi-\phi_1)]}}\\
   &&+R(\beta_3-\beta_2)+\int_{\beta_3}^{\phi_4}\frac{ d\phi}{\sqrt{\frac{2P_1}{B}[\sin(\phi_4-\phi_1)-\sin(\phi-\phi_1)]}},\\
  &&\frac{B}{2R^2}-P_0\sin\phi_2+P_1\sin(\phi_2-\phi_1)=-P_0\sin\phi_0,
 \end{eqnarray}
 \end{widetext}
where  $\beta_2+\beta_3=2\pi$.
As example,  Fig.~\ref{fig-SM}(d) illustrates the self-contact scenario, when the free end $s=L$ makes a point-contact with the front side of the strip.  Here we use $L/R=7$ in the numerical calculation.


\begin{thebibliography}{99}
\bibitem{note2}
The curvature of the the continuous contact region is constant, $\phi_s=\frac{1}{R}$, so $\vec{M}_s=0$. 
Then by considering moment balance of Eq.(1), it leads to $\hat{t}\times\vec{F}=0$, then we obtain $f_n=0$. 
Here $\vec{F}=f_t\hat{t}+f_n\hat{n}$ expressed in intrinsic coordinate, for a planar deformation. Here $\hat{n}$ is unit normal vector.
Furthermore, by considering force balance of Eq.(1) and taking $\vec{K}(s)=K\hat{n}$, we find that $f_t=constant$ and $K=-\frac{f_t}{R}= constant$ \cite{note3}.
The direction of $\hat{K}$ is changed with respect to $s$, while its magnitude is constant, $\vert \vec{K}\vert=constant$.
\bibitem{note1}
By moment balance of Eq.(1), we can obtain a relation for normal stress force, $f_n=-B\phi_{ss}$. Here the stress force we use is expressed in terms of intrinsic coordinate, $\vec{F}=f_t\hat{t}+f_n\hat{n}$.
Taking this result into force balance of Eq.(1), in the tangential direction, we find $(f_t)_s=f_n\phi_s=-B\phi_s\phi_{ss}$. By integration with respect to $s$, we have $f_t(s^{'})+\frac{1}{2}B\phi_s^2(s^{'})=constant.$
This is conservation relation for tangential stress force $f_t$ and curvature  $\phi_s$, valid for any location of the strip.
\bibitem{note3}
By using Frenet-Serret equations for a planar deformation, we have the following relations: $\frac{d\hat{t}}{ds}=\phi_s\hat{n} $ and $\frac{d\hat{n}}{ds}=-\phi_s\hat{t}$.
\end{thebibliography}
\end{document}